\documentclass{article}
\usepackage{spconf,amsmath,graphicx}

\def\x{{\mathbf x}}

\title{A study of audio mixing methods for piano transcription in violin-piano ensembles}

\name{Hyemi Kim$^\flat$$^\natural$  \qquad Jiyun Park$^\sharp$ \qquad Taegyun Kwon$^\sharp$ \qquad Dasaem Jeong$^\x$ \qquad Juhan Nam$^\flat$$^\sharp$ \thanks{
This research was supported by Culture, Sports and Tourism R\&D Program through the Korea Creative Content Agency grant funded by the Ministry of Culture, Sports and Tourism in 2022(Project Name: Development of high-speed music search technology using deep learning, Project Number: CR202104004, Contribution Rate: 50\%)
and Institute of Information \& communications Technology Planning \& Evaluation (IITP) grant funded by the Korea government(MSIT) 
(No.2019-0-00075, Artificial Intelligence Graduate School Program(KAIST)).
}
}
\address{$^\flat$Kim Jaechul Graduate School of AI, KAIST, Daejeon, South Korea \\
$^\natural$Electronics and Telecommunications Research Institute, Daejeon, South Korea \\
$^\sharp$Graduate School of Culture Technology, KAIST, Daejeon, South Korea \\
$^\x$Sogang University, Seoul, South Korea}
\begin{document}
%
\maketitle
\begin{abstract}
While piano music transcription models have shown high performance for solo piano recordings, their performance degrades when applied to ensemble recordings. This study aims to analyze the impact of different data augmentation methods on piano transcription performance, specifically focusing on mixing techniques applied to violin-piano ensembles. We apply mixing methods that consider both harmonic and temporal characteristics of the audio. To create datasets for this study, we generated the PFVN-synth dataset, which contains 7 hours of violin-piano ensemble audio by rendering MIDI files and corresponding labels, and also collected unaccompanied violin recordings and mixed them with the MAESTRO dataset. We evaluated the transcription results on both synthesized and real audio recordings datasets. 


\end{abstract}
\section{Introduction}
\label{sec:intro}

Automatic music transcription models have shown significant improvement in performance for single-instrument recordings, especially for piano, thanks to the development of carefully designed neural networks \cite{onsetsandframes, kong2021high, jongwook, kwon} and the availability of large-scale datasets such as the MAESTRO dataset \cite{maestro}. 
However, in multi-instrument settings, such as when the piano accompanies melodic instruments like the violin, cello, or flute, note detection for the piano becomes more challenging due to other instrumental sounds causing noise interference \cite{mt3, Wu}. This challenge has been addressed using multi-instrument datasets, such as MusicNet \cite{MusicNet} or synthesized Lakh dataset (Slakh) \cite{Slakh}. However, obtaining precise MIDI labels from mixed audio is an extremely difficult task. To address this issue, researchers have either aligned music scores to audio as noisy labels \cite{MusicNet} or synthesized the audio from MIDI files \cite{Slakh}.


Music source separation also faces a similar challenge where expensive multi-track audio recordings are required to train models. To overcome this problem, researchers have used various audio mixing methods as data augmentation during training. Among these methods, random mixing of audio segments has been the most commonly used. It involves incoherent mixing of audio segments from different songs at different positions \cite{Slakh} or using two segments from the same source as an augmented target source \cite{kong2021decoupling, catnet}. Some studies have introduced more organized random mixing based on acoustic knowledge, such as matching audio segments with chromagrams, assuming that mixing two audio segments with similar pitch content simulates real ensembles better than random mixing \cite{sinica, chromagram2022}. However, these audio mixing methods have not been extensively studied for automatic music transcription.

This paper presents the development of a piano transcription model for violin-piano ensembles. The model transcribes the piano part only, as an extension of the transcription model previously developed for solo piano recordings \cite{onsetsandframes}. To achieve this goal, we analyze various audio mixing methods for the two instrumental sounds in violin-piano ensembles. Unlike sound source separation, which mainly uses frame-level metrics, note onsets are particularly important in piano music transcription. Additionally, notes of the two instruments in ensembles are arranged tonally in Western classical music, which we use in this work. Based on these concepts, we propose an audio mixing approach that considers harmonic and temporal characteristics and compare it with random mixing. We also created a new dataset called PFVN-synth, containing 7 hours of violin-piano ensemble audio annotated with piano and violin MIDI labels. In addition, we collected unaccompanied violin recordings and mixed them with the MAESTRO piano dataset. As far as we know, this study is the first to explore data augmentation for piano transcription in ensemble music using harmonically and temporally controlled mixing methods.

\begin{figure}[t]
  \centering
  \centerline{\includegraphics[width=1.0\columnwidth]{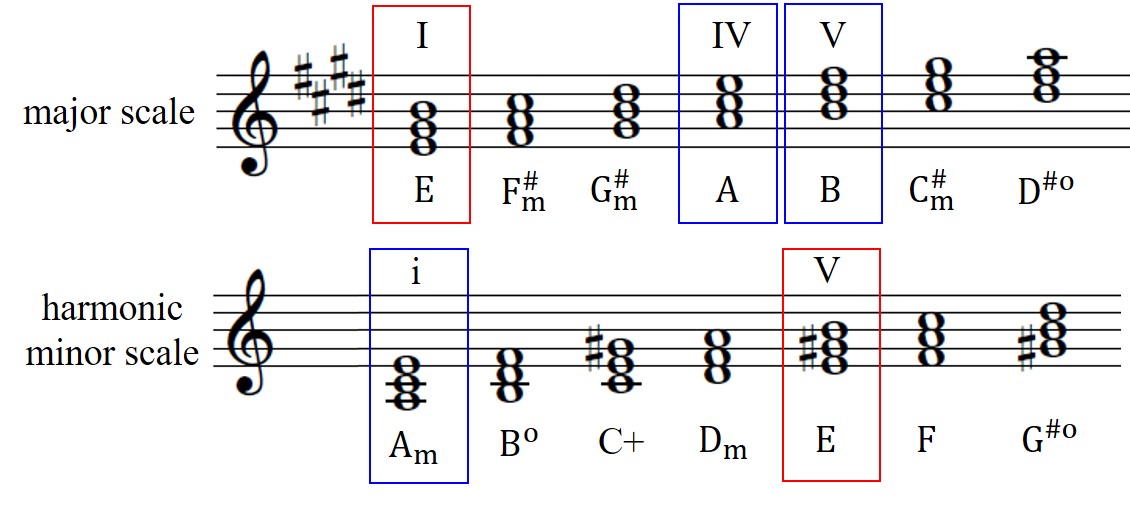}}  
\caption{An example of a major scale and a harmonic minor scale for the key-based mixing method. When the piano excerpt is in {\bf E} key (red boxes), the violin excerpt in {\bf B}, {\bf A}, or {\bf Am} (blue boxes) including {\bf E} is selected.}
\label{fig:keymethod}
\end{figure}

\section{Method for audio mixing}

\subsection{PFVN-synth Dataset}
\label{ssec:PFVN-synth}
The PFVN-synth dataset was created to contain realistic instrumental triplet audio recordings of the piano, violin, and their mixture\footnote{PFVN-synth is available at https://zenodo.org/record/7703620}. This was achieved by rendering MIDI files with virtual instruments using musical scores of 45 different pieces by 23 classical composers, with a total duration of 7 hours\footnote{We used commercial virtual instruments which are `Bösendorfer Grand Piano' in Apple Logic Pro and `SWAM Violin V3' by Audio Modeling.}. All MIDI files were collected from the MuseScore website, which provides not only sheet music but also corresponding MIDI files with dynamics and sustain pedals. All tracks were rendered into monaural audio files with the standard CD quality: 44.1kHz, 16-bit. The dataset includes recordings ranging from 58 seconds to almost 40 minutes in length. To ensure that composers are not biased to the split sets, it was divided into a train set containing 32 pieces with a duration of 5.8 hours, a validation set containing 3 pieces with a duration of 30 minutes, and a test set containing 10 pieces with a duration of 50 minutes.



\subsection{Real Performance Datasets}
\label{ssec:soloviolin}
We also trained the model with real performance recordings by artificially mixing violin and piano recordings from different sources.  
We collected unaccompanied violin pieces from the solo violin list\footnote{https://en.wikipedia.org/wiki/List\_of\_solo\_violin\_pieces}. To ensure diversity across eras, we selected recordings from 7 composers and downloaded them from YouTube. The total duration of the collected recordings is 4.6 hours. We then mixed the solo violin audio with piano audio from the MAESTRO dataset using various mixing methods, which are introduced in the following subsections. As a test set, we used a subset of the MusicNetEM dataset \cite{MusicNetEM} by selecting violin-piano ensemble pieces. This dataset has refined labels of the original MusicNet dataset \cite{MusicNet}. We excluded 7 noisy pieces and used a total of 14 pieces for testing, out of the 21 accompanied violin pieces in the dataset.



\subsection{Loudness Control}

The envelope of piano sounds exhibits a sudden increase of energy followed by a gradual decrease \cite{FMP}, whereas the envelope of violin sounds features a relatively soft increase followed by sustained energy. To mix the two instrument audios, we set the volume ratio based on the amplitude envelope. Specifically, we set the ratio of the RMS values of the piano and violin to 0.5 for the test and between 0.3 and 1.2 during the training. This ensured that the violin sound was not too loud and that the piano sound was not much louder than the violin. We also set the peak value of the mixed audio not to exceed 0.99.



\begin{figure}[!t]
  \centering
  \centerline{\includegraphics[width=0.75\columnwidth]{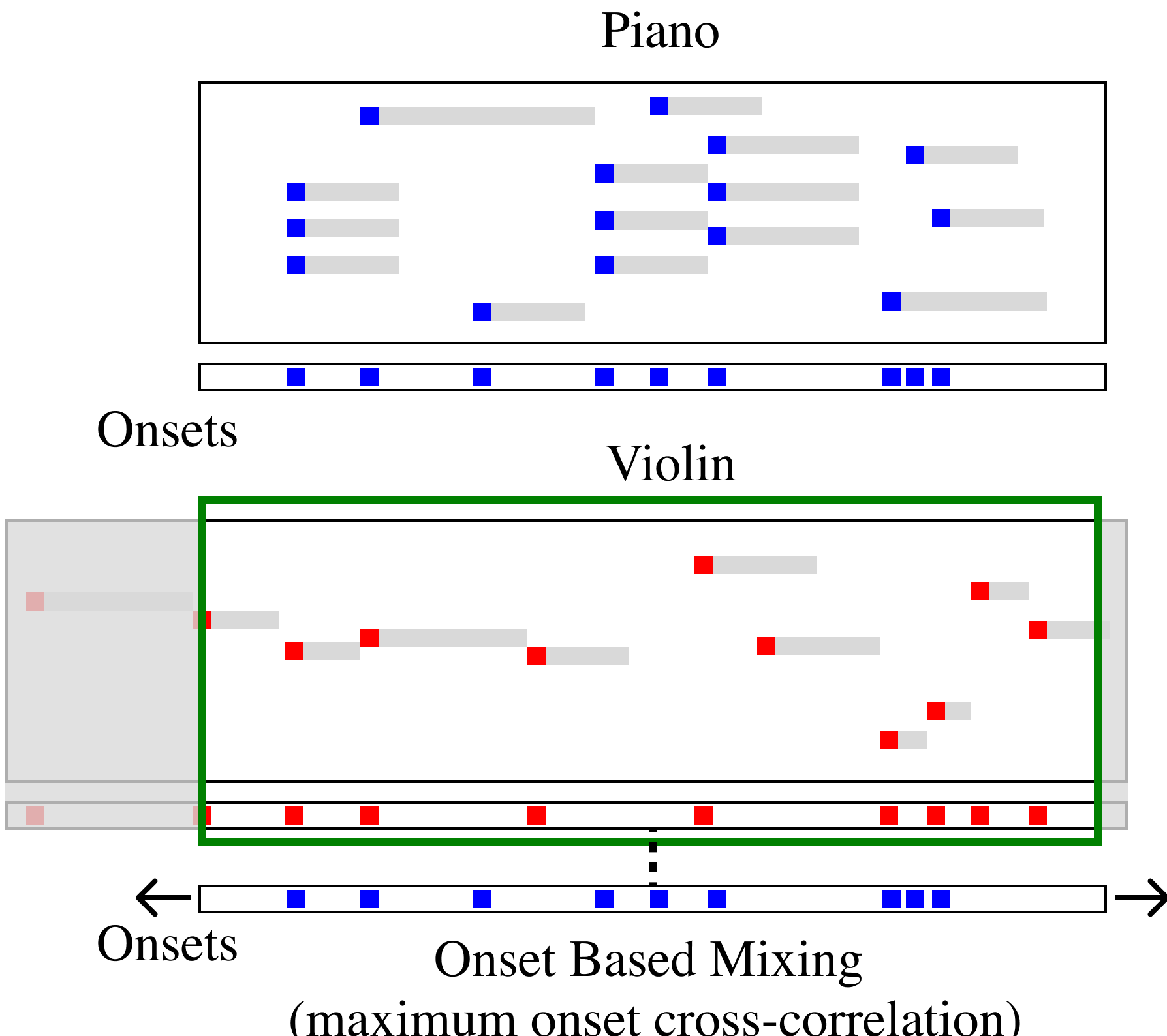} } 
\caption{An illustrated diagram of the onset-based mixing method. Since the MIDI file is not always available, the onsets are estimated and matched from the audio.} 
\label{fig:mixing}
\end{figure}

\subsection{Musically Controlled Mixing}
\label{subsec:mixing}

We used the random mixing technique as our baseline audio mixing method. In addition, we suggest two musically controlled mixing by matching tonal and temporal characteristics between the piano and violin audio excerpts. 


\begin{table*}[t]
\begin{center}	
\resizebox{0.9\textwidth}{!}{%
\begin{tabular}{l|ccc|ccc|ccc|ccc}

\hline

\hline
             &
  \multicolumn{3}{c|}{Frame} &
  \multicolumn{3}{c|}{Note with onset} &
  \multicolumn{3}{c|}{Note with offset} &
  \multicolumn{3}{c}{Note w. offset \& vel} \\ \hline
Mixing method  &
  \multicolumn{1}{c|}{P} &
  \multicolumn{1}{c|}{R} & F1
   &
  \multicolumn{1}{c|}{P} &
  \multicolumn{1}{c|}{R} & F1
   &
  \multicolumn{1}{c|}{P} &
  \multicolumn{1}{c|}{R} & F1
   &
  \multicolumn{1}{c|}{P} &
  \multicolumn{1}{c|}{R} & F1
   \\ \hline 
   
   \hline

Lower bound &	  \multicolumn{1}{c|}	{	82.0	}	&	  \multicolumn{1}{c|}	{	51.7	}	&	{	62.7	}	&	  \multicolumn{1}{c|}	{	78.3	}	&	  \multicolumn{1}{c|}	{	62.2	}	&	{	68.5	}	&	  \multicolumn{1}{c|}	{	51.9	}	&	  \multicolumn{1}{c|}	{	41.0	}	&	{	45.3	}	&	  \multicolumn{1}{c|}	{	40.8	}	&	  \multicolumn{1}{c|}	{	32.6	}	&	{	35.7	}	\\ \hline \hline
Original pair & 	  \multicolumn{1}{c|}	{	90.8	}	&	  \multicolumn{1}{c|}	{	81.8	}	&	{	85.7	}	&	  \multicolumn{1}{c|}	{	99.1	}	&	  \multicolumn{1}{c|}	{	96.8	}	&	{	97.9	}	&	  \multicolumn{1}{c|}	{	72.6	}	&	  \multicolumn{1}{c|}	{	71.0	}	&	{	71.8	}	&	  \multicolumn{1}{c|}	{	71.4	}	&	  \multicolumn{1}{c|}	{	69.8	}	&	{	70.6	}	\\ \hline
Random mixing & 	  \multicolumn{1}{c|}	{	95.4	}	&	  \multicolumn{1}{c|}	{	86.6	}	&	{	90.6	}	&	  \multicolumn{1}{c|}	{	99.2	}	&	  \multicolumn{1}{c|}	{	95.5	}	&	{	97.2	}	&	  \multicolumn{1}{c|}	{	80.2	}	&	  \multicolumn{1}{c|}	{	77.2	}	&	{	78.6	}	&	  \multicolumn{1}{c|}	{	78.9	}	&	  \multicolumn{1}{c|}	{	75.9	}	&	{	77.3	}	\\ \hline
Onset-based mixing & 	  \multicolumn{1}{c|}	{	96.0	}	&	  \multicolumn{1}{c|}	{	\textbf{88.9}}		&	{	\textbf{92.1}	}	&	  \multicolumn{1}{c|}	{	99.3	}	&	  \multicolumn{1}{c|}	{	97.1	}	&	{	\textbf{98.2}	}	&	  \multicolumn{1}{c|}	{	\textbf{82.3}	}	&	  \multicolumn{1}{c|}	{	\textbf{80.5}	}	&	{	\textbf{81.3}	}	&	  \multicolumn{1}{c|}	{	\textbf{80.8}	}	&	  \multicolumn{1}{c|}	{	\textbf{79.0}	}	&	{	\textbf{79.9}	}	\\ \hline
Key-based mixing & 	  \multicolumn{1}{c|}	{	\textbf{96.2}	}	&	  \multicolumn{1}{c|}	{	88.1	}	&	{	91.8	}	&	  \multicolumn{1}{c|}	{	\textbf{99.4}	}	&	  \multicolumn{1}{c|}	{	\textbf{97.2}	}	&	{	\textbf{98.2}	}	&	  \multicolumn{1}{c|}	{	81.7	}	&	  \multicolumn{1}{c|}	{	79.9	}	&	{	80.7	}	&	  \multicolumn{1}{c|}	{	80.2	}	&	  \multicolumn{1}{c|}	{	78.5	}	&	{	79.3	}	   \\ \hline
Key and onset-based mixing & 	  \multicolumn{1}{c|}	{	95.5	}	&	  \multicolumn{1}{c|}	{	87.9	}	&	{	91.4	}	&	  \multicolumn{1}{c|}	{	99.3	}	&	  \multicolumn{1}{c|}	{	96.1	}	&	{	97.6	}	&	  \multicolumn{1}{c|}	{	80.4	}	&	  \multicolumn{1}{c|}	{	77.8	}	&	{	79.1	}	&	  \multicolumn{1}{c|}	{	79.2	}	&	  \multicolumn{1}{c|}	{	76.7	}	&	{	77.9	}	   \\ \hline \hline
Upper bound &	  \multicolumn{1}{c|}	{	97.5	}	&	  \multicolumn{1}{c|}	{	90.9	}	&	{	94.0	}	&	  \multicolumn{1}{c|}	{	99.2	}	&	  \multicolumn{1}{c|}	{	97.8	}	&	{	98.5	}	&	  \multicolumn{1}{c|}	{	86.3	}	&	  \multicolumn{1}{c|}	{	85.1	}	&	{	85.7	}	&	  \multicolumn{1}{c|}	{	85.3	}	&	  \multicolumn{1}{c|}	{	84.2	}	&	{	84.7	}\\ \hline

\hline
\end{tabular}
}%
\end{center}
\caption{Transcription results from training on the PFVN-synth train split and evaluating on the test split.}
\label{tab:pfvn-train}
\end{table*}

\subsubsection{Key-based Mixing}
To ensure that the harmonic characteristics of the mixed piano and violin audios were similar, we mixed two audio excerpts that had similar tonality. We estimated the tonal key using the madmom library \cite{madmom}\footnote{We used \texttt{madmom.features.key} to estimate the key.}. To achieve this, we filtered the violin excerpts to ensure that they were in the same key as a given piano excerpt. We expanded the range of valid key combinations by adding the dominant key and subdominant key as harmonically-related keys. For example, in the case of a piano excerpt in the {\bf E} key, violin excerpts in the {\bf E}, {\bf B}, {\bf A}, and {\bf Am} keys were sampled, as shown in Figure \ref{fig:keymethod}. In the minor mode, the subdominant key of {\bf Em} is also {\bf Am}, so excerpts in keys such as {\bf Am}, {\bf E}, {\bf Dm}, and {\bf Em} were selected to mix with the target in {\bf Am}. By using these methods, we ensured that the randomly-selected two excerpts had similar tonality, which helped to create more realistic mixtures of piano and violin audios.

\subsubsection{Onset-based Mixing}
We also mixed the piano and violin audio excerpts in a way that aligned their note onsets as closely as possible. To achieve this, we shifted the entire piano clip by adjusting the global offset in the unit of time frame, as shown in Figure \ref{fig:mixing}. We aimed to find the optimal offset value that resulted in the highest degree of onset overlap between the two recordings by shifting the piano audio by one frame at a time, regardless of the tempo in the two audio excerpts. Then, we mixed the two audio excerpts at the optimal position. For synthesized audio, we used the corresponding MIDI files to find the note onset positions. For real audio, on the other hand, we estimated onsets using \emph{librosa} \cite{librosa}\footnote{We used \texttt{librosa.onset.onset\_detect} to detect onsets.}. By aligning the note onsets in this way, we improved the realism and accuracy of our mixed audio excerpts.

\section{Experiments}
\label{sec:experiments}

\subsection{Experiment Settings}
We conducted experiments to compare the performance of various audio mixing methods using \emph{Onsets and Frames}, a state-of-the-art piano transcription model \cite{onsetsandframes}. We based our implementation and hyperparameters on the source code available at this link\footnote{https://github.com/jongwook/onsets-and-frames}. We used a batch size of 16, and the maximum number of training iterations was set to 150,000 for the PFVN-synth train split and 500,000 for the real audio datasets. During validation, we selected the model with the highest note F1-score for the test. We used the standard frame-based and note-based metrics provided by \textit{mir\_eval} \cite{mir_eval}. To apply the onset-based mixing method, we set the length of the piano excerpt to about 20 seconds, and the violin excerpt was 5.5 seconds longer than the piano audio. 

In our experiments, we compared up to five audio mixing methods: 1) random mixing, 2) onset-based mixing, 3) key-based mixing, 4) key and onset-based mixing, and 5) the original mixing (only for PFVN-synth). We also compared the results to the lower-bound case, where we evaluated violin-piano ensembles using the model trained with piano only, and the upper-bound case, where we evaluated piano audio alone using the same model. By comparing the performance of these methods, we were able to identify the most effective approaches for piano transcription in ensemble music.

\section{Results and Discussion}
\label{sec:results}

\subsection{Models Trained with the PFVN-synth Dataset}
Table \ref{tab:pfvn-train} presents the piano transcription results using five different audio mixing methods. The lower-bound case, where the model is trained with piano only and tested with violin-piano ensembles, achieved a note F1-score of 68.5\%. After incorporating additional violin audio, note precision and recall increased significantly to around 99\% and 95\%, respectively. Onset-based mixing and key-based mixing produced the highest note F1-score of 98.2\%, which is close to the upper bound score of 98.5\%. In particular, onset-based mixing outperformed all four metrics of frame-level, note onset, note with offset, and note with offset and velocity, exhibiting the best overall performance. Key-based mixing, key and onset-based mixing, and random mixing followed onset-based mixing in descending order of performance. Data augmentation with random mixing failed to improve note onset F1-score, although it improved the scores of six metrics with note offset when compared to mixing with the original pair.


\begin{table*}[t]
\begin{center}	
\resizebox{0.9\textwidth}{!}{%
\begin{tabular}{l|ccc|ccc|ccc|ccc}
\hline

\hline
             &
  \multicolumn{3}{c|}{Frame} &
  \multicolumn{3}{c|}{Note with onset} &
  \multicolumn{3}{c|}{Note with offset} &
  \multicolumn{3}{c}{Note w. offset \& vel} \\ \hline
Mixing method  &
  \multicolumn{1}{c|}{P} &
  \multicolumn{1}{c|}{R} & F1
   &
  \multicolumn{1}{c|}{P} &
  \multicolumn{1}{c|}{R} & F1
   &
  \multicolumn{1}{c|}{P} &
  \multicolumn{1}{c|}{R} & F1
   &
  \multicolumn{1}{c|}{P} &
  \multicolumn{1}{c|}{R} & F1
   \\ \hline
   
   \hline

   \multicolumn{13}{c}{PFVN-synth} \\ \hline 
   
   \hline

Lower bound &	  \multicolumn{1}{c|}	{	71.8	}	&	  \multicolumn{1}{c|}	{	62.5	}	&	{	66.2	}	&	  \multicolumn{1}{c|}	{	63.5	}	&	  \multicolumn{1}{c|}	{	75.0	}	&	{	68.4	}	&	  \multicolumn{1}{c|}	{	36.4	}	&	  \multicolumn{1}{c|}	{	43.1	}	&	{	39.2	}	&	  \multicolumn{1}{c|}	{	26.3	}	&	  \multicolumn{1}{c|}	{	31.1	}	&	{	28.3	}	\\ \hline \hline
Random mixing & 	  \multicolumn{1}{c|}	{	82.3	}	&	  \multicolumn{1}{c|}	{	\textbf{75.4}	}	&	{	78.4	}	&	  \multicolumn{1}{c|}	{	\textbf{92.5}	}	&	  \multicolumn{1}{c|}	{	81.5	}	&	{	86.4	}	&	  \multicolumn{1}{c|}	{	57.6	}	&	  \multicolumn{1}{c|}	{	50.7	}	&	{	53.9	}	&	  \multicolumn{1}{c|}	{	45.6	}	&	  \multicolumn{1}{c|}	{	40.2	}	&	{	42.6	}	\\ \hline
Onset-based mixing &	  \multicolumn{1}{c|}	{	83.4	}	&	  \multicolumn{1}{c|}	{	73.9	}	&	{	78.1	}	&	  \multicolumn{1}{c|}	{	89.8	}	&	  \multicolumn{1}{c|}	{	81.1	}	&	{	85.0	}	&	  \multicolumn{1}{c|}	{	57.6	}	&	  \multicolumn{1}{c|}	{	52.0	}	&	{	54.5	}	&	  \multicolumn{1}{c|}	{	45.1	}	&	  \multicolumn{1}{c|}	{	40.9	}	&	{	42.7	}	\\ \hline
Key-based mixing & 	  \multicolumn{1}{c|}	{	\textbf{85.2}	}	&	  \multicolumn{1}{c|}	{	75.1	}	&	{	\textbf{79.6}	}	&	  \multicolumn{1}{c|}	{	92.3	}	&	  \multicolumn{1}{c|}	{	\textbf{83.1}	}	&	{	\textbf{87.3}	}	&	  \multicolumn{1}{c|}	{	\textbf{59.0}	}	&	  \multicolumn{1}{c|}	{	\textbf{53.0}	}	&	{	\textbf{55.7}	}	&	  \multicolumn{1}{c|}	{	\textbf{46.5}	}	&	  \multicolumn{1}{c|}	{	\textbf{41.9}	}	&	{	\textbf{44.0}	}	   \\ \hline
Key and Onset-based mixing &	  \multicolumn{1}{c|}	{	82.9	}	&	  \multicolumn{1}{c|}	{	74.5	}	&	{	78.2	}	&	  \multicolumn{1}{c|}	{	90.2	}	&	  \multicolumn{1}{c|}	{	81.0	}	&	{	85.1	}	&	  \multicolumn{1}{c|}	{	57.0	}	&	  \multicolumn{1}{c|}	{	51.1	}	&	{	53.8	}	&	  \multicolumn{1}{c|}	{	44.6	}	&	  \multicolumn{1}{c|}	{	40.2	}	&	{	42.2	}	   \\ \hline \hline
Upper bound &	  \multicolumn{1}{c|}	{	78.7	}	&	  \multicolumn{1}{c|}	{	87.7	}	&	{	82.7	}	&	  \multicolumn{1}{c|}	{	89.3	}	&	  \multicolumn{1}{c|}	{	93.0	}	&	{	91.0	}	&	  \multicolumn{1}{c|}	{	56.1	}	&	  \multicolumn{1}{c|}	{	58.6	}	&	{	57.3	}	&	  \multicolumn{1}{c|}	{	43.9	}	&	  \multicolumn{1}{c|}	{	45.9	}	&	{	44.8	}	   \\ \hline 

\hline \multicolumn{13}{c}{MusicNetEM} \\ \hline

\hline																											
Lower bound &	  \multicolumn{1}{c|}	{	75.1	}	&	  \multicolumn{1}{c|}	{	50.3	}	&	{	59.7	}	&	  \multicolumn{1}{c|}	{	71.8	}	&	  \multicolumn{1}{c|}	{	80.8	}	&	{	75.9	}	&	  \multicolumn{1}{c|}	{	28.9	}	&	  \multicolumn{1}{c|}	{	32.2	}	&	{	30.4	}	&	  \multicolumn{1}{c|}	{	-	}	&	  \multicolumn{1}{c|}	{	-	}	&	{	-	}	\\ \hline \hline
Random mixing & 	  \multicolumn{1}{c|}	{	78.0	}	&	  \multicolumn{1}{c|}	{	\textbf{58.2}	}	&	{	\textbf{65.8}	}	&	  \multicolumn{1}{c|}	{	\textbf{87.7}	}	&	  \multicolumn{1}{c|}	{	78.4	}	&	{	82.7	}	&	  \multicolumn{1}{c|}	{	36.7	}	&	  \multicolumn{1}{c|}	{	32.9	}	&	{	34.7	}	&	  \multicolumn{1}{c|}	{	-	}	&	  \multicolumn{1}{c|}	{	-	}	&	{	-	}	\\ \hline
Onset-based mixing &	  \multicolumn{1}{c|}	{	78.9	}	&	  \multicolumn{1}{c|}	{	57.4	}	&	{	65.6	}	&	  \multicolumn{1}{c|}	{	87.4	}	&	  \multicolumn{1}{c|}	{	78.4	}	&	{	82.4	}	&	  \multicolumn{1}{c|}	{	\textbf{37.1}	}	&	  \multicolumn{1}{c|}	{	33.4	}	&	{	35.0	}	&	  \multicolumn{1}{c|}	{-		}	&	  \multicolumn{1}{c|}	{	-	}	&	{	-	}	\\ \hline
Key-based mixing & 	  \multicolumn{1}{c|}	{	\textbf{79.2}	}	&	  \multicolumn{1}{c|}	{	56.8	}	&	{	65.5	}	&	  \multicolumn{1}{c|}	{	\textbf{87.7}	}	&	  \multicolumn{1}{c|}	{	\textbf{79.4}	}	&	{	\textbf{83.2}	}	&	  \multicolumn{1}{c|}	{	\textbf{37.1}	}	&	  \multicolumn{1}{c|}	{	33.7	}	&	{	\textbf{35.3}	}	&	  \multicolumn{1}{c|}	{	-	}	&	  \multicolumn{1}{c|}	{	-	}	&	{	-	}	   \\ \hline
Key and Onset-based mixing &	  \multicolumn{1}{c|}	{	\textbf{79.2}	}	&	  \multicolumn{1}{c|}	{	56.6	}	&	{	65.1	}	&	  \multicolumn{1}{c|}	{	87.1	}	&	  \multicolumn{1}{c|}	{	79.1	}	&	{	82.8	}	&	  \multicolumn{1}{c|}	{	37.0	}	&	  \multicolumn{1}{c|}	{	\textbf{33.8}	}	&	{	\textbf{35.3}	}	&	  \multicolumn{1}{c|}	{	-	}	&	  \multicolumn{1}{c|}	{	-	}	&	{	-	}	   \\ \hline 


\hline
\end{tabular}		
}%
\end {center}	
\vspace{-2mm}
\caption{Transcription results from training on the artificially mixed real performance dataset and evaluating on PFVN-synth and MusicNetEM.}
\label{tab:realaudio}
\end{table*}	

\subsection{Models Trained with the Real Performance Dataset}
Table \ref{tab:realaudio} presents the piano transcription results on four different audio mixing methods using two test datasets. The upper part of the table shows the results on the entire PFVN-synth dataset, including the training, validation, and test sets. After training with additional unaccompanied violin recordings, note precision significantly increases from 63.5\% to about 90\%, while note recall mildly increases from 75.0\% to over 81\%. The lower bound of the note F1-score is 68.4\%, and the upper bound is 91.0\%. All four mixing methods achieve more than 85\% of note F1-score. In this real performance case, the key-based mixing method achieves the highest F1-score of frame-level, note onset, note with offset, and note with offset and velocity. Onset-based mixing methods have relatively lower performance.

The lower part of Table \ref{tab:realaudio} shows the piano transcription results on the violin-piano ensemble pieces from MusicNetEM. After training with additional unaccompanied violin recordings, note precision greatly increases from 71.8\% to over 87\%, while note recall slightly decreases. The key-based mixing method achieves the highest note F1-score of 83.2\% and the note with offset F1-score of 35.3\%. The key and onset-based mixing method achieves the best note with offset F1-score along with the key-based mixing method.

\subsection{Analysis} 
\begin{figure}[!t]
  \centering
  \centerline{\includegraphics[width=8.5cm]{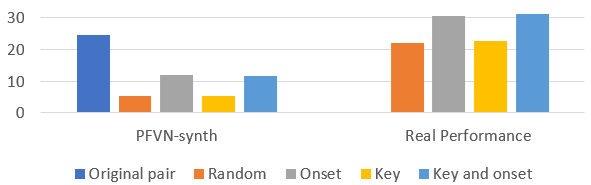}}  
\caption{
The number of overlapping note onsets between piano and violin audios in the two datasets.
}
\label{fig:onsetcount}
\end{figure}

Based on the results of Tables \ref{tab:pfvn-train} and \ref{tab:realaudio}, we observe that the performance of different audio mixing methods depends on the dataset used. The onset-based mixing method performs best on the PFVN-synth dataset while the key-based mixing method performs best on the real performance dataset. To investigate this difference, we counted the number of overlapping note onsets in the two datasets. Figure \ref{fig:onsetcount} shows that there are significantly more onset matches in the MAESTRO and unaccompanied violin recordings datasets compared to PFVN-synth, with two and a half to four times more matches. This can be attributed to the fact that the MAESTRO dataset was recorded at a piano competition with a higher number of onsets, and that unaccompanied violin recordings have more onset frames than violin onsets in PFVN-synth. We hypothesize that when the number of onsets is already high, the process of matching onsets does not significantly improve performance.

\section{Conclusions}

We conducted experiments to evaluate the performance of various audio mixing methods on piano transcription in violin-piano ensembles. We created a new synthesized dataset and collected unaccompanied violin recordings along with the MAESTRO dataset. The results indicate that the onset-based mixing method performed the best in our synthesized dataset, while the key-based mixing method was the most effective when the model was trained on real performance datasets. This suggests that musically-informed mixing is more effective than random mixing. In the future, we plan to collect real violin-piano ensemble recordings and further investigate mixing methods.

\vfill\pagebreak

\bibliographystyle{IEEEbib}
\bibliography{strings,refs}

\begin{thebibliography}{10}

\bibitem{onsetsandframes}
C.~Hawthorne, E.~Elsen, J.~Song, A.~Roberts, I.~Simon, C.~Raffel, J.~Engel,
  S.~Oore, and D.~Eck,
\newblock ``Onsets and frames: Dual-objective piano transcription,''
\newblock in {\em Proceedings of the 19th International Society for Music
  Information Retrieval Conference, Paris, France}, 2018, pp. 50--57.

\bibitem{kong2021high}
Q.~Kong, B.~Li, X.~Song, Y.~Wan, and Y.~Wang,
\newblock ``High-resolution piano transcription with pedals by regressing onset
  and offset times,''
\newblock {\em IEEE/ACM Transactions on Audio, Speech, and Language
  Processing}, vol. 29, pp. 3707--3717, 2021.

\bibitem{jongwook}
J.W. Kim and J.~P. Bello,
\newblock ``Adversarial learning for improved onsets and frames music
  transcription,''
\newblock in {\em Proceedings of the 21st International Society for Music
  Information Retrieval Conference}, 2019, pp. 670--677.

\bibitem{kwon}
T.~Kwon, D.~Jeong, and J.~Nam,
\newblock ``Polyphonic piano transcription using autoregressive multi-state
  note model,''
\newblock in {\em Proceedings of the 21st International Society for Music
  Information Retrieval Conference}, 2020, pp. 454--461.

\bibitem{maestro}
C.~Hawthorne, A.~Stasyuk, A.~Roberts, I.~Simon, C.~Huang, S.~Dieleman,
  E.~Elsen, J.~Engel, and D.~Eck,
\newblock ``Enabling factorized piano music modeling and generation with the
  maestro dataset,''
\newblock in {\em Proceedings of the 7th International Conference on Learning
  Representations}, 2019.

\bibitem{mt3}
J.~Gardner, I.~Simon, E.~Manilow, C.~Hawthorne, and J.~Engel,
\newblock ``\uppercase{MT3}: Multi-task multitrack music transcription,''
\newblock in {\em International Conference on Learning Representations}, 2022.

\bibitem{Wu}
Y.-T Wu, B.~Chen, and L.~Su,
\newblock ``Multi-instrument automatic music transcription with
  self-attention-based instance segmentation,''
\newblock {\em IEEE/ACM Transactions on Audio, Speech, and Language
  Processing}, vol. 28, pp. 2796--2809, 2020.

\bibitem{MusicNet}
J.~Thickstun, Z.~Harchaoui, and S.~Kakade,
\newblock ``Learning features of music from scratch,''
\newblock in {\em International Conference on Learning Representations}, 2017.

\bibitem{Slakh}
E.~Manilow, G.~Wichern, P.~Seetharaman, and J.~Roux,
\newblock ``Cutting music source separation some {Slakh}: A dataset to study
  the impact of training data quality and quantity,''
\newblock in {\em Proceedings of the IEEE Workshop on Applications of Signal
  Processing to Audio and Acoustics (WASPAA)}. IEEE, 2019.

\bibitem{kong2021decoupling}
Q.~Kong, Y.~Cao, H.~Liu, K.~Choi, and Y.~Wang,
\newblock ``Decoupling magnitude and phase estimation with deep resunet for
  music source separation,''
\newblock in {\em Proceedings of the 22nd International Society for Music
  Information Retrieval Conference}, 2021, pp. 342--349.

\bibitem{catnet}
X.~Song, Q.~Kong, X.~Du, and Y.~Wang,
\newblock ``Catnet: Music source separation system with mix-audio
  augmentation,''
\newblock {\em arXiv preprint arXiv:2102.09966}, 2021.

\bibitem{sinica}
C.~Y. Chiu, W.~Y. Hsiao, Y.~C. Yeh, Y.~H. Yang, and A.~W.~Y. Su,
\newblock ``Mixing-specific data augmentation techniques for improved blind
  violin/piano source separation,''
\newblock in {\em IEEE 22nd International Workshop on Multimedia Signal
  Processing}, 2020, pp. 1--6.

\bibitem{chromagram2022}
S.~Yuan, Z.~Wang, U.~Isik, R.~Giri, J.-M. Valin, M.~M Goodwin, and
  A.~Krishnaswamy,
\newblock ``Improved singing voice separation with chromagram-based pitch-aware
  remixing,''
\newblock in {\em Proceedings of the IEEE International Conference on
  Acoustics, Speech and Signal Processing}. IEEE, 2022, pp. 111--115.

\bibitem{MusicNetEM}
B.~Maman and A.~H Bermano,
\newblock ``Unaligned supervision for automatic music transcription in the
  wild,''
\newblock in {\em Proceedings of the 39th International Conference on Machine
  Learning}, 2022, pp. 14918--14934.

\bibitem{FMP}
M.~Müller,
\newblock {\em Fundamentals of Music Processing Audio, Analysis, Algorithms,
  Applications},
\newblock Springer, 2015.

\bibitem{madmom}
S.~B{\"o}ck, F.~Korzeniowski, J.~Schl{\"u}ter, F.~Krebs, and G.~Widmer,
\newblock ``madmom: a new python audio and music signal processing library,''
\newblock in {\em Proceedings of the 24th ACM International Conference on
  Multimedia}, Amsterdam, Netherlands, 2016, pp. 1174--1178.

\bibitem{librosa}
B.~McFee, C.~Raffel, D.~Liang, D.~P Ellis, M.~McVicar, E.~Battenberg, and
  O.~Nieto,
\newblock ``librosa: Audio and music signal analysis in python,''
\newblock in {\em Proceedings of the 14th python in science conference}, 2015,
  pp. 18--25.

\bibitem{mir_eval}
C.~Raffel, B.~McFee, E.~J. Humphrey, J.~Salamon, O.~Nieto, D.~Liang, and D.~PW
  Ellis,
\newblock ``mir\_eval: A transparent implementation of common {MIR} metrics,''
\newblock in {\em Proceedings of the 15th International Society for Music
  Information Retrieval Conference}, 2014, pp. 367--372.

\end{thebibliography}

\end{document}